\documentclass[12pt,a4paper]{article}
\usepackage{latexsym,epsf,epsfig}
\newcommand{\be}{\begin{equation}}
\newcommand{\ee}{\end{equation}}

\newcommand{\bea}{\begin{eqnarray}}
\newcommand{\eea}{\end{eqnarray}} 
\newcommand{\td}{\tilde}

\renewcommand{\k}{{\bf k}}

\def\bbox{\partial^2}

\newcommand{\ba}{\begin{eqnarray}}
\newcommand{\ea}{\end{eqnarray}}
\def\pc{\Phi}
\def\pv{\bar\Phi}
\def\bx{{\bf x}}
\def\by{{\bf y}}
\def\bk{{\bf k}}
\def\p{{\bf p}}
\def\bp{{\bf p}}
\def\bq{{\bf q}}
\def\w{\omega}

\begin{document}
\title{Effective theory
 for the soft fluctuation modes\\ 
in the spontaneously broken phase\\ 
of the N-component scalar field theory}

\author{
A. Jakov\'ac$^{a,}$\footnote{Zolt\'an Magyary fellow,
                             e-mail: jakovac@planck.phy.bme.hu}, 
A. Patk\'os$^{b,}$\footnote{e-mail: patkos@ludens.elte.hu},
P. Petreczky$^{b,}$\footnote{e-mail: petr@cleopatra.elte.hu} and 
Zs. Sz\'ep$^{b,}$\footnote{e-mail: szepzs@cleopatra.elte.hu}\\
\scriptsize (a) Department of Theoretical Physics,
  Technical University of Budapest, H-1521 Budapest, Hungary\\
\scriptsize (b) Department of Atomic Physics, E\"otv\"os University, H-1117
Budapest, Hungary } 
\maketitle

\abstract{The effective dynamics of the low-frequency modes is derived
for the $O(N)$ symmetric scalar field theory in the broken symmetry
phase.  The effect of the high-frequency fluctuations is taken into
account at one-loop level exactly. A new length scale is shown to
govern the long-time asymptotics of the linear response function of
the Goldstone modes. The large time asymptotic decay of an arbitrary
fluctuation is determined in the linear regime. We propose a set of
local equations for the numerical solution of the effective non-linear
dynamics. The applicability of the usual gradient expansion is
carefully assessed.}
  
\section{Introduction}
In the study of dynamical phenomena with the participation of
long-wavelength Higgs and Goldstone-modes one can account for the effect
of the high frequency modes most conveniently by deriving a number
of effective equations of motion. In a second step one might express
 the correction terms of the original equations with
help of a set of auxiliary fields, allowing local representation of
some nonlocal effects. A remarkable feature of this approach is that
for pure gauge theories a simple kinetic interpretation can be given
to the dynamics of the auxiliary fields
\cite{heinz83,blaizot99,bodeker98,arnold99,litim99}.

Recently we have shown that a similar interpretation is possible for
fluctuations of the one-component self-interacting scalar field in the
broken symmetry phase \cite{patkos99}. 
The evolution of the modes with wave number $|{\bf k}|$ was considered under 
the assumption $M<T$; that is the mass scale set by the spontanous
symmetry breakdown is smaller than the temperature.  Since 
$M\sim {\sqrt\lambda}\langle\Phi\rangle$, 
this condition relates the vacuum expectation value of the field to the
temperature and is actually met in systems with second order phase transition
not much below the transition point. This condition actually was used when 
the comparison of the quantum results with the classical dynamics was made.

The high frequency modes have a
non-trivial impact on the dynamics of the low frequency modes if the
scale characterizing the low-frequency fluctuations is much below the
mass scale of the symmetry breaking. In this sense one has to go
beyond the usual hard thermal loop approximation, since the effect of
loops with momentum $p\simeq M<T$ should be taken into account. The
classical statistical mechanical system proposed in \cite{patkos99}
reproduces the linear source-amplitude response computed in quantum
theory \cite{boyanovsky96a}.

In this paper we extend our discussion to the broken phase dynamics of
the N-component scalar field theory with $\Phi^4$ self-interaction.
This model is relevant to the dynamics of the $\pi -\sigma$-system
$(N=4)$ \cite{rajagopal93} and is actively investigated in connection
with the phenomenon of the disoriented chiral condensate
\cite{boyanovsky95,boyanovsky97,biro97, rischke98}.  The appearance,
the evolution and the damping of the low frequency fluctuations and
instabilities of the chiral order parameter are carefully studied in
these papers. Less attention is paid to the damping of the
Goldstone-modes. Though in the realistic case the explicit breaking of
the $O(N)$ symmetry leads to massive ``Goldstone''-bosons, it is of
interest to see what is the intrinsic dynamics of these excitations in
the ``ideal'' spontaneously broken case. It turns out that the damping of the
Goldstone-fluctuations with frequency $k_0$ is essentially different 
in the respective domains
\be
k_0<<{M^2\over T}<<M<T
\ee
and
\be
{M^2\over T}<<k_0<<M<T.
\ee
Our conclusion is that both
the on-shell dissipation time and the large time asymptotics of the
Goldstone fields is qualitatively different in the first region
in comparison to the
fluctuations of the order parameter.

The physical picture is very appealing. Radial fluctuations along the
order parameter relax fairly quickly, what freezes fast the length of
the vacuum expectation value of the $\Phi$ field. The second stage of
the relaxation consists of the slow rotation of the order parameter,
which is described by the relaxation of collectively excited low
momentum Goldstone fluctuation modes.
  
The high-temperature one-loop quantum dynamics of scalar models has been
investigated very actively recently partly as a kind of theoretical
laboratory for developing powerful calculational methods, partly with
the aim to provide answers to important questions of inflatory
cosmology and the physics of heavy ion collisions. Our work was
substantially influenced by References \cite{mrowczynski90,
boyanovsky96b,boyanovsky98,bodeker95,greiner97}.

In this paper we first derive in section 2 the effective equations for
the low frequency modes.  The coefficients of the terms correcting the
classical equations are determined by various $n$-point functions of
the fast modes.  For the effective dynamics we work out the
modification of the linear part of the equations (the self-energy
operator), which is fully determined by the two-point function of the
high-frequency fields. A detailed study of this quantity appears in
section 3. Stated in more technical terms, we compute one-loop
contributions to the two-point functions with full $T=0$ propagators
and restrict the temperature dependent contribution to the
$p_0>\Lambda$ modes. This approach follows the finite temperature
renormalization group transformation scheme of D'Attanasio and
Pietroni \cite{dattanasio96}.  In section 4 we shall discuss, in
particular, whether the linear response of the high frequency modes to
long wavelength fluctuations allows a classical kinetic theory
interpretation.  Using the results obtained for the two-point function
of the theory, in section 5 we present the explicit effective field
equations, with help of auxiliary fields introduced to handle the
nonlocal nature of the effective dynamics.  Our results are
summarized in section 6. In Appendix A some results of the main text
are rederived with help of the conventional perturbation theory in a
form explicitly showing its equivalence to the Dyson-Schwinger
treatment. Appendix B presents the iterative solution of the dynamics
of the classical $O(N)$-model. Some relevant integrals are explicitly
evaluated in Appendix C.

\section{The effective equations of motion for the slow modes}

The Lagrangian of the system is given by
\be
L={1\over 2}(\partial_\mu{\bf\varphi}_a)^2-{1\over 2}m^2({\bf\varphi}_a)^2-
{\lambda\over 24}(({\bf\varphi}_a)^2)^2.
\label{Lagr_dens}
\ee
The aim of this investigation is to integrate out the effect of $p_0>
\Lambda$ high frequency fluctuations. From the point of view of the
final result it turns out to be important whether $\Lambda <M$ or
$\Lambda >M$, where $M$ is the mass scale spontaneously generated in
the broken symmetry phase. In the spirit of the renormalization group
the value of $\Lambda$ will be lowered gradually and the importance of
passing the scale $M$ will become evident in this process. Our actual
interest will concentrate on the temporal variation of the lowest
frequency fluctuations $k_0<<\Lambda$.

We separate in the starting Lagrangian the high-frequency modes
($\phi_a(x)$ with frequency $\omega>\Lambda$) and the low-frequency modes
($\td\Phi_a (x)$ with frequency $\omega<\Lambda$)
\be
{\bf\varphi}_a(x)\rightarrow \td\Phi_a (x)+\phi_a (x).
\ee
Averaging over the high frequency fluctuations gives $\langle
\phi_a(x)\rangle=0$ which yields also $\td\Phi_a (x)=\langle
\varphi_a(x)\rangle$. This separation leads to
\bea
\label{shiftedL}
L&=&{1\over 2}(\partial\td\Phi_a)^2+{1\over 2}(\partial\phi_a)^2-
{m^2\over 2}\Bigl [(\td\Phi_a)^2+(\phi_a)^2\Bigr ]
-{\lambda\over 24}\Bigl [(\td\Phi_a)^2(\td\Phi_b)^2+
(\phi_a)^2(\phi_b)^2\Bigr ]\\\nonumber
&&-{\lambda\over 24}\Bigl [4(\td\Phi_a\phi_a)(\td\Phi_b\phi_b)+
4(\td\Phi_a\phi_a)(\phi_b)^2+4(\td\Phi_a\phi_a)(\td\Phi_b)^2+
2(\td\Phi_a)^2(\phi_b)^2\Bigr ].
\eea
From Eq.(\ref{shiftedL}) the equations for the slow modes can be
derived:
\be
(\bbox +m^2)\td\Phi_a+{\lambda\over 24}\bigl [8\phi_a(\td\Phi_b\phi_b)+
4\td\Phi_a(\td\Phi_b)^2+4\phi_a(\phi_b)^2+4\phi_a(\td\Phi_b)^2+
8\td\Phi_a(\td\Phi_b\phi_b)+4\td\Phi_a(\phi_b)^2\bigr ]=0.
\ee
The effect of the high-frequency fluctuations on the slow ones is
obtained by averaging the equations with respect to their
statistics. At one-loop level accuracy $\langle
\phi_a\phi_b\phi_c\rangle=0$ and one arrives at
\be
(\bbox +m^2)\td\Phi_a
+{\lambda\over 6}\td\Phi_a(\td\Phi_b)^2+
{\lambda\over 3}\td\Phi_b\langle\phi_a\phi_b\rangle+
{\lambda\over 6}\td\Phi_a\langle(\phi_b)^2\rangle =0.
\ee
We introduce at this point the conventional notation:
\be
\Delta_{ab}(x,y)\equiv\langle\varphi (x)\varphi (y)\rangle .
\ee
Below no summation will be understood when repeated indices appear without
the explicit summation symbol. We shall also use specific pieces extracted
from the above two-point functions defined as
\bea
&
\Delta^{(0)}_{ab}(x,y)\equiv\Delta_{ab}(x,y)|_{\tilde\Phi =0},
\nonumber\\
&
\Delta^{(1)}_{ab}(x,y)\equiv\int dz{\delta\Delta_{ab}(x,y)\over
\delta\tilde\Phi_c (z)}|_{\tilde\Phi =0}\cdot\tilde\Phi_c(z).
\eea

In the broken symmetry phase one has to separate the nonzero average
value from the slowly varying field, and write the equation only for
the fluctuating part:
\be
\td\Phi_a\rightarrow\bar\Phi\delta_{a1}+\Phi_a,
\label{breaking}
\ee
where we have chosen the direction of the average to point along the
$a=1$ direction. We shall analyze the resulting equations for the
$a=1$ and the remaining $a=i\neq 1$ components separately.  Since the
main effect of the high frequency modes we are interested in is the
modification of the mass term (self-energy contribution), therefore we
shall restrict our study to the linearized equation of $\Phi_a(x)$.
One has to take into account that the two-point functions
$\langle\phi_a\phi_b\rangle$ depend on the background $\Phi_a(x)$. For
the linearized equations it is sufficient to compute them only up to
terms linear in the background.

The two equations are 
\bea
&&
(\bbox +{\lambda\over 3}\bar\Phi^2)\Phi_1(x) + \frac{\lambda\bar\Phi}2
\Phi_1^2(x) + \frac{\lambda\bar\Phi}6 \sum\limits_{i\neq1} \Phi_i^2(x) +
\frac\lambda6 \Phi_1(x) \sum\limits_a\Phi_a^2(x) + J_1(x)=0,\nonumber\\
&&
\bbox \Phi_i(x) + \frac{\lambda\bar\Phi}3 \Phi_i(x)\Phi_1(x) +
\frac\lambda6 \Phi_i(x) \sum\limits_a\Phi_a^2(x) + J_i(x)=0,\qquad (i\neq1)
\label{effeqs}
\eea
with the induced currents
\bea
&&
J_1(x) = {\lambda\over 2} \bar\Phi\Delta^{(1)}_{11}(x,x) + {\lambda\over
  6}(N-1)\bar\Phi\Delta^{(1)}_{ii}(x,x),\nonumber\\
&&
J_i(x) = {\lambda\over 3}(\Delta^{(0)}_{ii}(x,x)-\Delta^{(0)}_{11}(x,x))
\Phi_i+{\lambda\over 3}\bar\Phi\Delta^{(1)}_{i1}(x,x).
\label{j_ind}
\eea
 We have simplified the
equations using the fact $\Delta_{i,b\neq i}$ has no
$\Phi$-independent piece, since with zero background the correlators
are diagonal. The equations (\ref{j_ind}) express the linear response
of the fast modes to the presence of a low frequency background
producing an effective source term to their classical dynamical
equations.
Also one has not to forget that $\bar\Phi$ is now the solution of the
constant part of the equations:
\be
m^2\bar\Phi+{\lambda\over 6}(3\Delta^{(0)}_{11}(x,x)+(N-1)\Delta^{(0)}_{ii}
(x,x))\bar\Phi+{\lambda\over 6}\bar\Phi^3=0.
\label{avphi}
\ee

\section{The two-point function of the fast modes}
\label{sec:2pointfunction}

Varying Eq.(\ref{shiftedL}) with respect to $\phi (x)$ one arrives at
the equations of motion of the fast modes in the background of
$\td\Phi (x)$:
\be
(\bbox +m^2)\phi_a+{\lambda\over 24}\bigl (8\td\Phi_a(\td\Phi_b\phi_b)+
4\phi_a(\phi_b)^2+4\td\Phi_a(\td\Phi_b)^2+8(\td\Phi_b\phi_b)\phi_a
+4\td\Phi_a(\phi_b)^2+4\phi_a(\td\Phi_b)^2\bigr )=0.
\ee
The equations of motion linearized in the high frequency fields
can be written in the form
\be
\biggl\{(\bbox+m^2)\delta_{ab}+{\lambda\over 6}\bigl [2\td\Phi_a\td\Phi_b+
(\td\Phi_c)^2\delta_{ab}\bigr ]\biggr\}\phi_b=
-{\lambda\over 6}\td\Phi_a(\td\Phi_b)^2.
\label{lineq}
\ee
We introduce the notation
\be
m_{ab}^2(x)=m^2\delta_{ab}+{\lambda\over 6}[2\td\Phi_a(x)\td\Phi_b(x)+
(\td\Phi_c(x))^2\delta_{ab}]
\label{massmatr}
\ee
and apply Eqs.(\ref{lineq}) and (\ref{massmatr}) to the fields appearing
in the definition of the two-point function $\Delta_{ac}(x,y)$:
\bea
&&
\bbox_x \Delta_{ac}(x,y)=-m^2_{ab}(x)\Delta_{bc}(x,y),\nonumber\\
&&
\bbox_y \Delta_{ac}(x,y)=-m^2_{cb}(y)\Delta_{ab}(x,y).
\eea
In the derivation of these homogeneous equations we have exploited that
$\langle\td\Phi_a(x)(\td\Phi_c(x))^2\phi_b(y)\rangle\linebreak =0$.
After the Wigner-transformation
\be
\Delta (X,p)=\int d^4ue^{ipu}\Delta (x,y),\qquad u=x-y
\ee
one arrives at the exact linear equations for the Wigner
transforms. (The use of the Latin letters $k,p,q$ is reserved for the 
4-momenta and $pk, kq$, etc. denotes their Minkovskian scalar products.) 
For the two-point functions diagonal in $O(N)$ indices one
finds in the broken phase to linear order in the background (after
using Eq.(\ref{breaking}))
\bea
&&
\bigl [{1\over 4}\bbox_X-ip\cdot\partial_X-p^2+ M^2_a]\Delta_{aa}(X,p)+
\lambda_a\bar\Phi\int{d^4q\over (2\pi)^4}\Phi_1(q)\Delta_{aa}(X,
p-{q\over 2})e^{-iqX}=0,\nonumber\\
&&
\bigl [{1\over 4}\bbox_X+ip\cdot\partial_X-p^2+M^2_a]\Delta_{aa}(X,p)+
\lambda_a\bar\Phi\int{d^4q\over (2\pi)^4}\Phi_1(q)\Delta_{aa}
(X,p+{q\over 2})e^{-iqX}=0,
\label{MDeq}
\eea
where for the Goldstone bosons on has
\be
M_{i}^2=0,\qquad \lambda_i={\lambda\over 3},
\ee
while for the heavy (``Higgs'') mode
\be
M^2_1={\lambda\over 3}\bar\Phi^2,\qquad \lambda_1=\lambda.
\ee
Here the tree level mass relations were used, as they correspond to the
actual order of the perturbation theory. This procedure  can be improved 
using Eq.(\ref{avphi}) for $\bar\Phi$, what is equivalent to the
resummation of the perturbative series.

After performing the Fourier-transformation also with respect to the
center-of-mass coordinate and subtracting the resulting two equations
one arrives at
\bea
&&
2pk\Delta_{11}(k,p)=-\lambda\bar\Phi\int{d^4q\over (2\pi )^4}\Phi_1(q)
\bigl [\Delta_{11}(k-q,p+{q\over 2})-\Delta_{11}(k-q,p-{q\over 2})\bigr ],
\nonumber\\
&&
2pk\Delta_{ii}(k,p)=-{\lambda\over 3}\bar\Phi\int{d^4q\over (2\pi )^4}
\Phi_1(q)
\bigl [\Delta_{ii}(k-q,p+{q\over 2})-\Delta_{ii}(k-q,p-{q\over 2})\bigr ].
\label{diageq}
\eea

For the mixed two-point function $\Delta_{i1}$ in the same steps a
similar equation is derived:
\be
(2pk+{\lambda\over 3}\bar\Phi^2)\Delta_{i1}(k,p)=-{\lambda\over 3}\bar\Phi
\int{d^4q\over (2\pi )^4}\bigl [\Delta_{ii}(k-q,p+{q\over 2})-
\Delta_{11}(k-q,p-{q\over 2})\bigr ]\Phi_i(k).
\label{offdiageq}
\ee

In the kinematical region, where the center-of-mass variation is very
slow one can assume $q\approx k$. (If there would be no $X$-dependence
one would find $\Delta(k-q,P)\sim \delta (k-q)$). If in addition one
restricts the variation of $p$ in the high frequency fields by the
relation $|k_\mu |<<\Lambda \simeq |p_\mu |$, one can approximate the integrals
in these equations by a factorized form. If the functions
$\Delta_{aa}(k,p-{k \over 2})$ are replaced by the first two terms of
their power series with respect to $k/2$, one recovers the drift
equations proposed originally by Mr\'owczynski and Danielewicz
\cite{mrowczynski90}. In case of the diagonal two-point functions
these equations are identical to the collisionless Boltzmann equation
for a gas of scalar particles, whose masses
$(M_1^2+\lambda\bar\Phi\Phi_1(x)$ and $M_i^2+
\lambda\bar\Phi\Phi_1(x)/3$) are determined by the $\Phi_1$ field:
\be 
p\cdot\partial_X\Delta_{aa}(X,p)+{\lambda_a\over
  2}\bar\Phi\partial_X\Phi_1 \partial_p\Delta_{aa}(X,p)=0.
\label{kineq}
\ee
Below we shall discuss an alternative to this expansion, which does
not exploit the assumption of slow $X$-variation. At the end we will
be in position to assess the validity of the assumption which led to
the Boltzmannian kinetic equations (\ref{kineq}).

At weak coupling one can attempt the recursive solution of Eqs.
(\ref{diageq}) and (\ref{offdiageq}). The starting point of the
iteration is the assumption that {\it the high frequency modes are
close to thermal equilibrium}, therefore one has for the starting
2-point functions \cite{lebellac95}
\bea
&&
\Delta_{ii}^{(0)}(p)=2\pi\delta (p^2)(\Theta (p_0)+\tilde n(|p_0|)),
\nonumber\\
&&
\Delta_{11}^{(0)}(p)=2\pi\delta (p^2-M_1^2)(\Theta (p_0)+\tilde n(|p_0|)),
\nonumber\\
&&
\tilde n(x)=n(x)\Theta (x-\Lambda )={1\over e^{\beta x}-1}\Theta (x-\Lambda).
\label{propagator}
\eea
Since the starting distributions are independent of $X$, the integrals
in Eq. (\ref{MDeq}) factorize exactly and one has for the first
corrections the following explicit expressions:
\bea
&&
\Delta^{(1)}_{aa}(k,p)=-{1\over 2pk}\lambda_a\bar\Phi\Phi_1(k)
\bigl [\Delta_{aa}^{(0)}(p+{k\over 2})-\Delta_{aa}^{(0)}(p-{k\over 2})\bigr ],
\nonumber\\
&&
\Delta_{i1}^{(1)}(k,p)=-{1\over 2pk+M_1^2}{\lambda\over 3}\bar\Phi
\Phi_i(k)\bigl [\Delta^{(0)}_{ii}(p+{k\over 2})-\Delta^{(0)}_{11}(p
-{k\over 2})\bigr ].
\label{corr1eq}
\eea
The retarded nature (ie. forward time evolution) will be taken into
account with the Landau prescription $k_0\to k_0+i\epsilon$
(see Appendix \ref{app:pertth} for the equivalent formulas in
the conventional perturbation theory).

\section{Non-equilibrium linear dynamics of the slow modes}

The quantum improved, linearized equations of motion are
(see eq.~(\ref{effeqs}))
\begin{equation}
  (\bbox + M_a^2)\Phi_a(x) + J_a(x) =0.
\label{lineq1}
\end{equation}
The (retarded) self-energy function is introduced through the
 relation
\begin{equation}
  J_a(x) = \int\!d^4y\,\Pi_a(x-y)\,\Phi_a(y),
\end{equation}
where, after Fourier-transformation one obtains, (see eq.~(\ref{corr1eq}))
\begin{eqnarray}
& \Pi_1(k) = \displaystyle\int\frac{d^4p}{(2\pi)^4}\Biggl[ &\!\! \left(
  \frac{\lambda\bar\Phi} 2 \right)^2  \frac1{pk}
  \left(\Delta_{11}^{(0)}(p-\frac k2) - \Delta_{11}^{(0)}(p+\frac k2)
  \right) \nonumber\\
&& + \left(\frac{\lambda\bar\Phi} 6 \right)^2 (N-1) \frac1{pk}
  \left(\Delta_{ii}^{(0)}(p-\frac k2) - \Delta_{ii}^{(0)
}(p+\frac k2)
  \right)\Biggr], \nonumber\\
& \Pi_i(k) = \displaystyle\int\frac{d^4p}{(2\pi)^4}\Biggl[ &\!\! \left(
  \frac{\lambda\bar\Phi} 3 \right)^2  \frac1{2pk + M_1^2}
  \left(\Delta_{11}^{(0)}(p-\frac k2) - \Delta_{ii}^{(0)}(p+\frac k2)
  \right)  \nonumber\\
&& - \frac\lambda 3 \left(\Delta_{11}^{(0)}(p) - \Delta_{ii}^{(0)}(p)
  \right)\Biggr].
\label{pis}
\end{eqnarray}
These integrals can be calculated (or at least reduced to 1D
integrals). Some details of the calculations can be found in Appendix
~\ref{sec:appcalc}, here we just state the results:
\begin{eqnarray}
&& \Pi_1(k) = \left[ \left(\frac{\lambda\bar\Phi} 2 \right)^2\,
  R_1(k,M_1) + \left(\frac{\lambda\bar\Phi} 6 \right)^2 (N-1)
  \,R_1(k,0) \right],\nonumber\\
&& \Pi_i(k) = \left(\frac{\lambda\bar\Phi}3\right)^2 R_i(k,M_1).
\label{pis1}
\end{eqnarray}
The explicit form of their real parts is the following:
\begin{eqnarray}
& \textrm{Re} R_1(k,M)=& \int\limits_M^\infty\!d\omega\, (1+2\tilde
n(\omega))\,\left[{\cal A}(\frac{2|\p||\k|}{2\omega k
_0+k^2}) + {\cal
    A}(\frac{2|\p||\k|}{-2\omega k_0+k^2}) \right], \nonumber\\
& \textrm{Re} R_i(k,M)=& \int\limits_M^\infty\!d\omega\, \left[ (1+\tilde
  n(\omega))\,{\cal A}(\frac{2|\p||\k|}{2\omega k_0+k^2+M^2}) + \tilde
    n(\omega) {\cal A}(\frac{2|\p||\k|}{-2\omega k_0+k^2+M^2}) \right] 
  \nonumber\\ 
&&\hspace{-1.5cm} +\int\limits_0^\infty\!d|\p|\, \left[ (1 + \tilde n(|\p|))\, 
{\cal
    A}(\frac{2|\p||\k|}{-2|\p| k_0+ k^2- M^2}) + \tilde n(|\p|) {\cal
    A}(\frac{2|\p||\k|}{2 |\p| k_0+k^2-M^2}) \right],
\label{repis}
\end{eqnarray}
where $|\p|^2=\omega^2-M^2$ and
\begin{equation}
  {\cal A}(x)=\frac1{4\pi^2|\k|}\textrm{arth}(x) = \frac1{8\pi^2|\k|} \ln
  \left|\frac{1+x}{1-x}\right|.
\end{equation}
The expressions of the imaginary parts look as follows:
\begin{eqnarray}
& \textrm{Im} R_1(k,M)=&\frac{-1}{4\pi|\k|}\Biggl[\Theta(-k^2)\!\!\!\!\!
  \int\limits_{P_c-k_0}^{P_c}\!\!\!\!\!dP\,\tilde n(P) \,+\,
  \Theta(k^2\!-\!4M^2)\int\limits_{k_0/2}^{P_c} \!dP\,(1 + \tilde
  n(k_0\!-\!P) + \tilde n(P))\Biggr],
  \nonumber\\ 
& \textrm{Im} R_i(k,M)=&\displaystyle{\frac{-1}{16\pi|\k|}} \Biggl[\,
  \Theta(M^2-k^2)\left[ \int\limits_{Q_+}^{Q_++k_0}\!\!\!  dP\,\tilde
  n(P) + \int\limits_{|Q_--k_0|}^{|Q_-|}\!\!\! dP\,\tilde n(P) \right]\,
  \nonumber\\
&& + \Theta(k^2-M^2)\int\limits_{Q_+}^{Q_-} \!dP\,(1 + \tilde n(k_0 - P)
  + \tilde n(P))\Biggr],
\label{impis}
\end{eqnarray}
where
\begin{equation}
  P_c=\frac {|\k|}2\sqrt{1-\frac{4M^2}{k^2}} + \frac{k_0}2,\quad
  Q_+=\left|\frac{k^2-M^2}{2(k_0+|\k|)}\right|, \quad 
  Q_-=\frac{k^2-M^2}{2(k_0-|\k|)}.
\end{equation}

The real part is divergent at zero temperature which cancels against
the coupling constant counterterm contribution. Care has to be taken,
however, when implementing the regularization, to maintain the Lorentz
invariance for the pieces containing $T=0$ parts of the propagators
$\Delta^{(0)}$, which is manifest in the original form in
Eq.~(\ref{pis}).

It is remarkable that the usual domain of Landau damping ($|\k
|^2>|k_0|^2$) is apparently extended up to $M_1^2+|\k |^2>|k_0|^2$. An
analogous situation has been noted and interpreted recently for the
propagation of a light fermion in heavy scalar plasma
\cite{boyanovsky98b}. Below we find for the damping of soft on-shell
Goldstone-modes the same interpretation.
 
The linearized equations of motion can be analyzed from several points
of view. One is the determination of the dispersion relations. This
describes the quasiparticles, and physically corresponds to the
``dressing'' of an (external) particle passing through the thermal
medium. The other point of view is the field evolution
\cite{boyanovsky96b}, when one follows the solution of the field
equations developing from given initial conditions (history).

\subsection{Dispersion relations for on-shell waves}

The position of the poles is determined by the equation
\begin{equation}
  k^2-M_a^2 -\Pi_a(k) = 0.
\end{equation}
We split $k_0$ into real and imaginary parts: $k_0=\omega-i\Gamma$,
and assume $\Gamma\ll\omega$. Then the perturbative solution can be
written as
\be
  \omega^2 = \omega_0^2 + \textrm{Re}\Pi(\omega_0,\k),\qquad
  \Gamma =  - \frac{\textrm{Im}\Pi(\omega_0,\k)}{2\omega_0},
\ee
where $\omega_0^2=M_a^2+\k^2$. The corresponding time dependence for
fixed  wave vector \k\ and given initial amplitudes ($\partial_t\Phi
(t=0,\k )=P({\bf k}),\Phi (t=0,\k )=F({\bf k})$) is found
\begin{equation}
  \Phi(t,\k)= \biggl[P(\k) \frac{\sin\omega t}{\omega} + F(\k)
  \cos\omega t \biggr]\,e^{-\Gamma t}.
\label{disprel}
\end{equation}

\paragraph{Goldstone modes.}

The tree level dispersion relation has no mass gap. The second
equation of Eq. (\ref{pis}) shows that $\Pi_i(k=0)=0$, that is no mass gap
is created radiatively neither at finite temperature; this is the
manifestation of the Goldstone theorem. The on-shell imaginary part,
as Eq.~(\ref{pis1}) shows, comes from the continuation of the
Landau-damping extending up to $k^2<M_1^2$. Going back to Appendix C
(Eq.(\ref{gh_diff})), one finds that the imaginary part receives
contribution from the collision of a hard thermal Goldstone particle
with distribution $n(p_0)$ with the soft external Goldstone wave of
momentum $k$ producing a hard Higgs-particle minus the inverse
reaction, when a hard thermal Higgs with momentum distribution
$n(p_0+k_0)$ decays into a soft and a hard Goldstone. The two
contributions can be combined into a single integral leading to the
following expression for the damping rate:
\begin{equation}
  \Gamma_i(\k) = \left(\frac{\lambda\bar\Phi}3\right)^2\,\frac1{32\pi
  \k^2} \int\limits_{M_1^2/4|\k|}^{M_1^2/4|\k| + |\k|}\!\! dp\,\tilde n(p).
\end{equation}
If $|\k |\ll M_1$ we can write with a good approximation
\begin{equation}
 \Gamma_i(\k) = \left(\frac{\lambda\bar\Phi}3\right)^2\!\frac1{32\pi
  |\k |} \,\tilde n(\frac{M_1^2}{4|\k |}).
\label{goldstone_damping}
\end{equation}
This contribution survives the IR cut, if $\Lambda<M_1^2/4|\k|$. 

The result (\ref{goldstone_damping}) is interesting from several
points of view. First, in HTL approximation, where we neglect all the
masses, we would found $\Gamma_i \sim \tilde n(0) =0$. On the other
hand the classical approximation corresponds to the substitution
$n(x)\to T/x$ in Eq.(\ref{goldstone_damping}), which results 
(see Appendix \ref{calss}) in
\begin{equation}
  \Gamma_i^{{\rm class}} = \frac{\lambda T}{24\pi}.
\label{goldstone_class}
\end{equation}
However, for very small momenta ($|\k |<<M_1^2/4T$) the correct result
is exponentially small, deviating considerably from the classical
result. The purely classical simulations therefore cannot reproduce
the correct Goldstone dynamics in the long wavelength region, they
would overestimate the damping rate.

The most important consequence of the form of the Goldstone damping is
the generation of a new dynamical length scale $M_2^{-1}=4T/M_1^2$.
The components of the Goldstone condensate with longer wave-length can
not (exponentially) decay, they survive for a longer time.

\paragraph{Higgs modes.}

From Eq.~(\ref{pis}) one can easily read off the on-shell
($k_0^2=\k^2+M_1^2$) value of the imaginary part of the
self-energy. This is entirely due to Higgs scattering into a soft +
hard Goldstone-pair, and leads to
\begin{equation}
  \Gamma_1=\frac{-\textrm{Im} \Pi_1}{2k_0} =
  \frac{\lambda M_1^2}{96\pi k_0|\k|} (N-1) 
  \int\limits_0^{|\k|/2} \!dP\,(1 + \tilde n(\frac{k_0}2 - P) + \tilde
  n(\frac{k_0}2 + P)).
\end{equation}
In the $k\ll M_1$ limit it simplifies to
\begin{equation}
  \Gamma_1= \frac{\lambda M_1}{96\pi} (N-1) \left(\frac12 +\tilde
  n(\frac{M_1}2) \right).
\end{equation}
This survives the cut if $2\Lambda<M_1$. 

Since $M_1<T$ we can perform a high temperature expansion. 
This means at the same time that the classical approximation is applicable. 
We find
\begin{equation}
  \Gamma_1 = \frac{\lambda T}{48\pi} (N-1),
\end{equation}
which is $(N-1)/2\,\Gamma_i^{{\rm class}}$.

\subsection{Solution of the initial value problem of the slow fields}

Equation(\ref{lineq1}) describes an integro-differential equation. For its
solution we have to know the complete past history of the fields. The
previous subsection provides a specific way to look at the problem. In
general we can introduce $z_a(x)=\int_{y_0<0}\!d^4y\,\Pi_a(x-y)\,\Phi_
a(y)$, then for $x_0>0$
\begin{equation}
  \Phi_a(x)= \int\frac{d^4k}{(2\pi)^4}\,e^{-ikx}\,\frac{z_a(k)}
  {k^2-M^2-\Pi_a(k)}.
\label{init_problem}
\end{equation}
Since all singularities are in the lower $k_0$ half plane, $\Phi_a(x)$
defined by this relation vanishes for $x_0\equiv t<0$.  $z(x)$ can be
used for setting the initial conditions \cite{boyanovsky96b}, as it
can describe jumps in the field as well as in its time derivative. For
example $\Phi_a(x)=0$ for $x_0<0$ and $\Phi_a(t=0)=F_a$, $\partial_t
\Phi_a(t=0)=P_a$ corresponds to
\begin{equation}
  z_a(x) = -P_a({\bf x}) \delta(x_0) - F_a({\bf x})
  \delta^{\prime}(x_0), \qquad z_a(k)= ik_0 F_a({\bf k}) - P_a({\bf
    k}).
\end{equation}
Direct substitution of Eq. (\ref{init_problem}) into Eq. (\ref{lineq1})
shows that it satisfies the effective homogenous wave equation. 
By evaluating the $k_0$ integral for $t=0$ (for details, see below) one 
also can demonstrate that it fulfills the initial conditions set above.

In general, we are faced with the computation of Fourier transforms of
functions analytic on the upper complex plane. The procedure of
extracting their large time asymptotics was investigated already in
Refs.~\cite{boyanovsky96a, boyanovsky96b}.  For $t>0$ the $k_0$
integration contour can be closed with an infinite semi-circle in the
lower half-plane and we pick up the contribution of the cuts and poles
inside the closed integration path:
\begin{equation}
  f(t)= \int\limits_{-\infty}^\infty \!\frac{dk_0}{2\pi}\, f(k_0)\,
  e^{-ik_0 t} = \sum\limits_{\omega\in\textrm{poles}}\!\!\!\!
  (-iZ(\omega)) e^{-i\omega t} + \sum\limits_{I\in\textrm{cuts}}
  \int\limits_I\!  \frac{dk_0}{2\pi i}\, \rho(k_0)\, e^{-ik_0 t},
\end{equation}
where $Z(\omega)$ is the residuum of the physical poles of the $f$-function
\footnote{The poles below a cut do not contribute to this formula,
they lie on the unphysical Riemann sheet.}, $\rho=i\,\textrm{Disc} f$
is the discontinuity along the cut, $I=[\omega_1,\omega_2]$ is the
support of the cut.  The second term can be computed again by
completing the original integration interval to a closed contour. The
discontinuity itself may have poles (but no cuts!) which contribute in
the same way as the ``normal'' poles. After these poles are
``encircled'', there are two straight contours - parallel to the
imaginary axis - left, starting at the two ends of the cut and running
in the interval $[\omega,\omega-i\infty]$, where $\omega$ is the end
(or starting) point (threshold). After this analysis one arrives at
the generic form
\begin{equation}
  f(t)= \sum\limits_{\omega\in\textrm{poles}}\!\!\!\! (-iZ(\omega))
  e^{-i\omega t} + \!\!\!\! \sum\limits_{\omega\in\textrm{thresh.}}
  \!\!\!\! (\mp) e^{-i \omega t} \int\limits_0^\infty \!
  \frac{dy}{2\pi}\, \rho(\omega -iy)\, e^{-yt},
\end{equation}
where the $-$ sign is to be applied for the starting, the $+$ sign for
the end point of the cut, and we have to take into account all poles,
also the ones on the unphysical Riemann sheet. Expanding $\rho$ around
$\omega$ into power series of $y$, the $y$ integration can be
performed.  The term $\sim y^n$ of the expansion contributes to the
time dependence $t^{-n-1}$. The large time behavior is {\em dominated
by the lowest power term of the expansion}. If $\rho$ cannot be power
expanded then the damping for large times is faster than a power-law.

In our case the position of the poles is determined by the dispersion 
relations (see previous section), and for given initial values we find
\begin{equation}
  \Phi_{pole}(t,\k)= Z(\k)\, \biggl[P(\k) \frac{\sin\omega
    t}{\omega} + F(\k) \cos\omega t \biggr]\,e^{-\Gamma t}.
\end{equation}
This expression coincides with eq.~(\ref{disprel}) apart from the wave
function renormalization $Z(\k )$. The time dependence of the
quasi-particle pole contribution was analyzed before.

For the cuts the factorized form can be used for the spectral
function: $\rho_a = z_a \rho^G_a$, where $\rho^G_a$ is the spectral
function of the propagator:
\begin{equation}
  \rho^G_a = \frac{-2\textrm{Im}\Pi_a} { (k^2-M^2 -\textrm{Re} \Pi_a)^2
    + (\textrm{Im}\Pi_a)^2}.
\end{equation}
The thresholds and the leading large-time behavior are determined by
$\textrm{Im}\Pi$, Eq.~(\ref{impis}). The imaginary part of the
Goldstone self energy is non-analytic only at $k^2=0$, but also there
the non-analytic piece vanishes as $\sim\exp(-M^2/(2T|k_0-|\k||))$. This
finally leads to a damping faster than any power which can be seen after
a saddle point analysis
\begin{equation}
  \Phi_{cut,i} \sim \frac{ |\k|}{\lambda M_1^2} \left(\frac{\beta
      M_1^2}{t^3}\right)^{1/4} \, e^{-2\sqrt{\beta M_1^2t}}\, \left[ P(\k)
    \frac{\sin |\k|t }{|\k|} + F(\k) \cos |\k|t \right].
\end{equation}
The cut contribution for the Higgs damping, on the other hand, is
similar to the zero temperature case
\begin{eqnarray}
  &\Phi_{cut,1}(t,\k)= &-\frac{\lambda T}{24 M_1}\, \frac1{(\pi M_1
  t)^{3/2}}\,\left[ P(\k) \frac{\sin (2M_1 t-\pi/4) }{2M_1} + F(\k) \cos
  (2M_1 t -\pi/4) \right]  \nonumber \\
  && + \frac{\lambda |\k|}{96\pi M_1}\, \frac1{\pi M_1 t}\,\left[ P(\k)
  \frac{\sin (|\k|t+\pi/2)}{|\k|} + F(\k) \cos(|\k|t +\pi/2) \right].
\end{eqnarray}
The first term comes from the threshold of the Higgs pair production,
the second from the threshold of the Goldstone pair production or their
Landau damping. The Landau damping of the Higgs particles does not 
contribute to the power law decay.

The terms decaying as some powers of time will dominate the time
evolution after the period of the exponential decay for the Higgs
bosons
($|\k|^{-1}\ll M_1^{-1}$). Similar is the case for small Goldstone
domains ($|\k|^{-1}\ll M_2^{-1}$), there the exponential decay is
followed by a $\sim\exp(-\sqrt{t})$ time evolution. In case of large
Goldstone domains ($|\k|^{-1}\gg M_2^{-1}$), however, because of the
exponentially small damping rate, the situation is reversed: the
$\sim\exp(-\sqrt{t})$ behavior will be dominant for intermediate
times, while the amplitude is reduced by a factor of $Z^{-1}$. Only
for very long times will become the exponential damping term, arising
from the Goldstone-pole, relevant. Its action will erase completely
the large size domains.

\section{Nonlinear dynamics}

The calculation of the effective non-linear evolution of the
low-frequency modes can be based on Eq.(\ref{effeqs}) using the
explicit expressions for the Fourier transforms of the induced
currents defined in Eq.(\ref{j_ind}):
\bea
J_1(k) & = & {\lambda^2\bar\Phi^2\over 4}\int{d^4p\over (2\pi )^4}{1\over pk}
\Phi_1(k)\bigl [\Delta_{11}^{(0)}(p-k/2)-\Delta_{11}^{(0)}(p+k/2)\bigr ]
\nonumber\\
&&
+(N-1){\lambda^2\bar\Phi^2\over 36}\int{d^4p\over (2\pi )^4}{1\over pk}
\Phi_1(k)\bigl [\Delta_{ii}^{(0)}(p-k/2)-\Delta_{ii}^{(0)}(p+k/2)\bigr ],
\nonumber\\
J_i(k) & = & -{\lambda\over 3}\int{d^4p\over (2\pi )^4}{2pk\over 2pk+M_1^2}
\Phi_i(k)\bigl [\Delta_{11}^{(0)}(p-k/2)-\Delta_{ii}^{(0)}(p+
k/2)\bigr ].
\label{expl_source}
\eea

Implicitly in all Wigner-transforms $\Delta^{(0)}$ the Bose-Einstein
factor is understood with a low frequency cutoff, well separating
the modes treated classically from the almost thermalised high
frequency part of the fluctuation spectra.

For the purpose of numerical investigations the above non-local form
of the induced currents is difficult to use. In this section we
discuss the introduction of auxiliary fields making the numerical
solution of the nonlinear dynamics easier to implement.  We shall not
take into account the nonlinear piece of the source induced at
one-loop level, since in weak coupling the leading nonlinear effect
comes from the tree-level cubic term. The consistent inclusion of the
higher power induced sources will not lead to qualitatively new,
leading effects unlike the linear source, which is responsible for
damping. We leave this extension for future investigations.

We concentrate first on the induced current $J_1$. By appropriate
shifts of the integration variable $p$ it can be rewritten as
\bea
J_1(k) & = & {\lambda^2\bar\Phi^2\over 2}\int {d^4p\over (2\pi )^4}\bigl [
{1\over 2pk+k^2}-{1\over 2pk-k^2}\bigr ]\Delta_{11}^{(0)}(p)\Phi_1(k)
\nonumber\\
&&
+(N-1){\lambda^2\bar\Phi^2\over 18}\int{d^4p\over (2\pi )^4}\bigl [
{1\over 2pk+k^2}-{1\over 2pk-k^2}\bigr ]\Delta_{ii}^{(0)}(p)\Phi_1(k).
\eea
After explicitly performing the $p_0$ integration one arrives at the
following expression which has a clear interpretation as a specific
statistical average:
\bea
J_1(k) & = & \Phi_1(k)\Bigl [{\lambda^2\bar\Phi^2\over 2}\int{d^3p\over
(2\pi )^3}{1\over 2p_0}(1+2\tilde n(p_0))({1\over 2pk+k^2}-
{1\over 2pk-k^2})|_{p_0=({\bf p}^2+M_1^2)^{1/2}}\nonumber\\
&&
+(N-1){\lambda^2\bar\Phi^2\over 18}\int{d^3p\over (2\pi )^3}{1\over 2p_0}
(1+2\tilde n(p_0))({1\over 2pk+k^2}-{1\over 2pk-k^2})|_{p_0=|{\bf p}|}\Bigr ].
\label{statav}
\eea
The temperature independent piece corresponds to the $T=0$
renormalization of $\lambda$, fixed at the scale $k^2$. It is absorbed
into the mass term of $\Phi_1$, therefore we retain in the integrand
only the terms proportional to the cutoff Bose-Einstein factors.
 
We introduce two complex auxiliary ``on-shell'' fields $W^a(x,{\bf
p})$, $a=1,i$. They correspond to the two different mass-shell
conditions appearing in the above equation, and fulfill the equations:
\be
(2pk-k^2)W^a(k,{\bf p})=\Phi_1(k).
\label{auxeq}
\ee
Clearly, $J_1(x)$ can be expressed as a well defined combination of
the thermal averages of these fields:
\be
-J_1(x)=\lambda^2
\bar\Phi^2\Bigl (\langle W^1(x,{\bf p})\rangle
+\langle W^1(x,{\bf p})^{*}\rangle 
+{N-1\over 9}(\langle W^i(x,{\bf p})\rangle+
\langle W^i(x,{\bf p})^{*}\rangle )\Bigr ).
\label{j1_ind}
\ee
Thermal averages are defined by the usual formula 
\be
\langle W^a(x,{\bf p})\rangle =\int{d^3p\over (2\pi )^3}{1\over 2p_0}\tilde
n(p_0)W^a(x,{\bf p})|_{p_0=({\bf p}^2+M_a^2)^{1/2}},\qquad a=1,i.
\ee
For the relevant combination one can use in place of Eq. (\ref{auxeq}) the
equations arising after the combination $W^a(x,{\bf p})-W^{a}(x,{\bf p})^*$ is
eliminated:
\be
[(2pk)^2-(k^2)^2](W^a(k,{\bf p})+W^a(-k,{\bf p})^*)=2k^2\Phi_1(k).
\label{auxsymeq}
\ee
Equations (\ref{effeqs}), ({\ref{j1_ind}) and (\ref{auxsymeq})
represent that form of the nonlinear dynamics which is best adapted
for numerical solution.

For very small values of $k_0$ a simplified form can be derived which
coincides with the result of the conventional kinetic treatment of the
Higgs-modes \cite{patkos99}.  One arrives at this approximate
expression of $J_1(k)$ if one performs an appropriate three-momentum shift
${\bf p}\rightarrow {\bf p\mp k}/2$ on the ${\bf p}$ variable in Eq.
(\ref{statav}).  One finds the following expressions for the
denominators, accurate to linear order in $k_0$:
\bea
&
2pk\pm k^2\rightarrow 2k_0\sqrt{({\bf p\mp k}/2)^2+M_a^2}-2{\bf pk}
\pm k_0^2\approx 2pk(1\pm k_0/2p_0)+{\cal O}(k_0^3,{\bf k}^3),
\nonumber\\
&
p_0\rightarrow p_0\mp {{\bf pk}\over 2p_0}.
\eea
Performing the expansion of the denominators and of the arguments of
$\tilde n (p_0)/p_0$ to linear order in ${\bf k}, k_0$ one finds the
following approximate expression for $J_1$:
\bea
& 
J_1(k)=-{\lambda^2\bar\Phi^2\over 2}\int {d^3p\over (2\pi )^3}
\Bigl [\Bigl ({1\over 2p_0^2}({d\tilde n(p_0)\over dp_0} {k_0p_0\over pk}
+{1\over p_0}
({1\over 2}+\tilde n(p_0)-p_0{d\tilde n(p_0)\over dp_0})\Bigr )|
_{p_0=({\bf p}^2+M_1^2)^{1/2}}\nonumber\\
&
+{N-1\over 18}{1\over p_0^2}\Bigl ({d\tilde n(p_0)\over dp_0} {k_0p_0\over pk}
+{1\over p_0}
({1\over 2}+\tilde n(p_0)-p_0{d\tilde n(p_0)\over dp_0})\Bigr )|
_{p_0=|{\bf p}|}\Bigr ]
\Phi_1(k).
\eea
The IR-cutoff $\Lambda >>|\k|$ is important to prevent the singularity
of the contribution from the Goldstone-modes.
\footnote{ The proper solution of the effective dynamics of the heavy
field surrounded by massless Goldstone-quanta will require the
application of methods analogous to the Bloch-Nordsieck resummation at
$T=0$ \cite{boyanovsky98c}.  We thank M. Simionato for an enlightening
discussion on this point.}  The contributions from the second terms in each line
are independent of $k$, therefore they are simply understood as a finite
shift in the squared mass of $\Phi_1$.  With these steps one arrives at the
final expression for the nonlocal part of the induced ``Higgs'' current:
\be
J_1(k)= -{\lambda^2\bar\Phi^2\over 4}\int{d^3p\over (2\pi )^3}
\Bigl [{1\over p_0^2}{d\tilde n(p_0)\over dp_0}|_{p_0=({\bf p}^2+
M_1^2)^{1/2}}
+{N-1\over 9}{1\over p_0^2}
{d\tilde n(p_0)\over dp_0}|_{p_0=|{\bf p}|}\Bigr ]{1\over vk}k_0\Phi_1(k).
\label{j2_ind}
\ee 
Here $v^\mu =(1,{\bf p}/p_0)$.  In this case it is sufficient to
introduce only two real auxiliary fields $W^1(x,{\bf v})$ and
$W^i(x,{\bf v})$ in order to make the dynamical equations local. They
are defined through the equations:
\be
vkW^a(k,{\bf v})=k_0\Phi_1(k), \qquad a=1,i.
\label{aux1eq}
\ee
The weighting factors in the integrals over the auxiliary field can be
interpreted as the deviations from equilibrium of the Higgs and
Goldstone particle distributions, $\delta n_1({\bf p})$ and $\delta
n_i({\bf p})$, due the scattering of the high frequency particles off
the $\Phi_1$ condensate. Then one writes
\be
J_1(x)=-{\lambda^2\bar\Phi^2\over 4}\Bigl (\delta\langle W^1(x,{\bf v})
\rangle +{N-1\over 9}\delta\langle W^i(x,{\bf v})\rangle\Bigl ),
\ee
where $\delta\langle ... \rangle$ means a phase space integral
weighted with the nonequilibrium part of the relevant distributions.

For this case it is easy to construct the energy-momentum vector
corresponding to the nonlocal piece of the induced current. One can follow 
the procedure
proposed by Blaizot and Iancu \cite{blaizot94} and investigate the
divergence of the $( \mu ,0)$ component of the energy-momentum tensor:
\be
\partial_\mu T^{\mu 0}_{induced}=-J_1\partial^0\Phi_1,
\ee
The task is to transform the right hand side into the form of a
divergence.  Using Eqs.(\ref{j2_ind}) and (\ref{aux1eq}) one finds the
following expression:
\bea
&
-J_1\partial^0\Phi_1={\lambda^2\bar\Phi^2\over 4}\int{d^3p\over (2\pi )^3}
\Bigl (W^1(x,{\bf v}){1\over p_0^2}{d\tilde n(p_0)\over dp_0}
(v\partial_x)W^1(x,{\bf v})|_{p_0=({\bf p}^2+M_1^2)^{1/2}}\nonumber\\
&
+{N-1\over 9}W^i(x,{\bf v}){1\over p_0^2}{d\tilde n(p_0)
\over dp_0}(v\partial_x)W^i(x,{\bf v})|_{p_0=|{\bf p}|}\Bigr ).
\eea
From here one can read off the  induced energy-momentum function of
the approximate Higgs dynamics:
\bea
T^{\mu 0}_{induced} & = &  {\lambda^2\bar\Phi^2\over 8}
\int{d^3p\over (2\pi )^3}
v^\mu\Bigl [{1\over p_0^2}{d\tilde n(p_0)\over dp_0}
W^1(x,{\bf v})^2|_{p_0=({\bf p}^2+M_1^2)^{1/2}}\nonumber\\
&& 
+ {N-1\over 9}{1\over p_0^2}{d\tilde n(p_0)\over dp_0}
W^i(x,{\bf v})^2|_{p_0=|{\bf p}|}\Bigr ].
\eea

For the exact (essentially non-local) dynamics we did not attempt the
construction of an energy functional, but its existence for the above
limiting case hints for the Hamiltonian nature also of the full one-loop
dynamics as expressed in terms of the auxiliary variables.

The analysis of the induced Goldstone source, appearing in
Eq.(\ref{expl_source}) goes in fully analogous steps. The temperature
dependent part to which a statistical interpretation can be linked is
rewritten with help of two complex auxiliary fields $V^a(x,{\bf p}),
a=1,i$ as
\be 
J_i(x)={\lambda\over 3} \int{d^3p\over (2\pi )^3}
\Bigl [{1\over p_0}\tilde n(p_0=({\bf p}^2+M_1^2)^{1/2})
\textrm{Re}\, V^1(x,{\bf p})-{1\over p_0}\tilde n(p_0=|{\bf p}|) 
\textrm{Re}\,V^i(x,{\bf p})\Bigr ],
\label{j2_i}
\ee
where the auxiliary fields fulfill the equations
\bea
&
(2pk+k^2+M_1^2)V^1(k,{\bf p})=(2pk+k^2)\Phi_i(k),\nonumber\\
&
(2pk-k^2+M_1^2)V^i(k,{\bf p})=(2pk-k^2)\Phi_i(k).
\label{Veq}
\eea
In this case even for very small values of $k_0$ we were not able to
derive any limiting case in which one could treat the time evolution
of the low frequency Goldstone modes as a truly Boltzmannian kinetic
evolution.

The solution of the system of equations (\ref{effeqs}),
(\ref{j1_ind}), (\ref{auxsymeq}), (\ref{j2_i}), (\ref{Veq}) requires
the specification of initial conditions also for the auxiliary fields
$W^a+W^{a*}, V^a$. If one uses the "past history" condition for the
physical fields $\Phi_1(x)={\rm const}, \Phi_i(x)= {\rm const}$ for
$t<0$, then the linear integral equation form of the equations for the
auxiliary fields and their retarded nature imply vanishing $W^a, V^a$
for $t=0$.

\section{Conclusions}

In this paper we have performed a complete 1-loop analysis of the time
evolution of low frequency ($k_0<<M_1$) field configurations in the
broken phase of the $O(N)$ symmetric scalar field theory.  We have
shown explicitly the full formal equivalence of the leading order
iterative solution of the Dyson-Schwinger equations and the
perturbation theory computation of the two-point function.

We have analyzed explicitly the case when $T>M_1$, which occurs, for
instance, in the vicinity of the second order phase transition of the
model. Here we could compare our results with the results arising from
the dynamical equations of the classical $O(N)$ field theory
with appropriately chosen parameters. For the "Higgs" particle we have
found complete agreement for the $|\k|<<M_1<T$ modes.
Related to this is the fact that we were able to show that the exact
one-loop equations derived for the auxiliary fields $W^a$ have a simple
Boltzmannian collisionless kinetic form for small $|\k|$.

However, the analysis gave different conclusions for the linear
response of the Goldstone-modes. The agreement with the classical
theory is restricted to the interval $M_1^2/4T<<|\k|<<M_1$. Below the new
characteristic scale
\be
M_2\equiv {M_1^2\over 4T}
\label{newscale}
\ee
the damping of the Goldstone modes becomes exponentially small and it
vanishes non-analytically for $|\k|\rightarrow 0$. This is a very natural
manifestation of the Goldstone-theorem in a dynamical situation: no
homogeneous ground state will relax to any "rotated" nearby
configuration.  In the light of this suggestive picture it is not
surprising that we could not find a classical kinetic interpretation
of the exact one-loop dynamics of the Goldstone modes even for very
small values of $|\k|$.

Besides the exponential damping analyzed above, $(1-Z)$ fraction of the
initial configuration follows a different time evolution determined
by the particle production and Landau-damping cuts. In case of the
lowest wave number Higgs fluctuations this leads to the result that the
exponential regime will be followed by a power-decay for times
$t>M_1^{-1}$. In case of the Goldstone modes the cut contribution is
$\sim\exp{(-4\sqrt{M_2t})}$, and for modes above the scale $M_2$ this
will dominate for large times. Below the scale $M_2$ the cut
contribution will be observable for intermediate times, while the
pole dominated damping, because of the exponentially small damping rate, 
becomes relevant only for very long times.

It will be interesting to see through numerical investigations, if the
onset of the nonlinear regime will influence the damping scenario of
the Goldstone fluctuations. The effect of interaction among the
high-$k$ modes (in addition to their scattering off the low-$k$
background) on the effective theory merits also further study.
Finally, we work on the extension of our analysis to the Gauge+Higgs
models in the broken phase, relevant to the physics of the standard
model below the electroweak phase transition.

\begin{appendix}

\section{Appendix}
\label{app:pertth}

Instead of the generalized Boltzmann-equations described in
Section~\ref{sec:2pointfunction} we can use also perturbation theory
to evaluate the two-point correlation functions. Here we will
demonstrate the equivalence of the one-loop perturbation theory and the
iterative solution of the generalized Boltzmann-equations in the case
of $\Delta^{(1)}$.

In the perturbation theory we write
\[ \left<\phi_a(x)\phi_b(x)\right> =\frac 1Z\,\left<\textrm{T}_c\,
  \phi^{(0)}_a(x)\phi^{(0)}_b(x)\,e^{-i{\cal S}_I}\right>,\]
where $\phi^{(0)}$ are the free fields, ${\cal S}_I$ is the 
part of the action which decribes the interaction between the different
field components in the presence of the background. $\textrm{T}_c$ stands
for the time ordering along a complex time path $c$ specified in
\cite{lebellac95}. At one loop, to linear order in $\Phi$ we need only
\[ S_I=\frac{\lambda\bar\Phi}6\,\int_c dx_{0c}\int d^3x
\left[ \Phi_1\,(\phi_b)^2+ 2 (\Phi_b\phi_b) \phi_1 \right],\] 
where the integration variable $x_{0c}$ represents the points on the
complex integration contour in the $t$ plane.  The field operator
contractions are performed with help of the matrix propagators
\begin{eqnarray}
iG_{ab}(x)=\left(\begin{array}[c]{cc} 
	iG^C_{ab}(x)\quad & iG^<_{ab}(x) \cr
	iG^>_{ab}(x)\quad & iG^A_{ab}(x) \cr\end{array}\right) =
\left(\begin{array}[c]{cc} 
   \langle\textrm{T}\phi_a(x)\phi_b(0)\rangle &
   \langle\phi_b(0)\phi_a(x)\rangle \cr
   \langle\phi_a(x)\phi_b(0)\rangle &
   \langle\textrm{T}^*\phi_a(x)\phi_b(0)\rangle \cr
\end{array}\right),
\end{eqnarray}
where T$^*$ denotes anti time ordering. The tree level propagators
are diagonal $G_{ab}(x)=\delta_{ab} G_a(x)$. Since the background
depends on the real (not the contour) time we find
\begin{equation}
  \left<\phi_a(x)\phi_b(x)\right>^{(1)} \equiv
  -i\left<\phi^{(0)}_a(x)\phi^{(0)}_b(x)\,{\cal S}_I\right> =
  \frac{\lambda\bar\Phi}3 \int\!d^4y\, [\Phi_1(y)\delta_{ab} +
  \Phi_a(y)\delta_{b1} + \Phi_b(y)\delta_{a1}] S_{ab}(x-y),
  \label{corr6eq}
\end{equation}
where
\begin{equation}
  iS_{ab}(z) = G_a^C(z)G_b^C(z) - G_a^<(z) G_b^<(z).
\end{equation}
The propagators can be expressed with help of the spectral function
$\rho$ as 
\begin{equation}
G^<(p)=n(p_0)\rho(p),\qquad G^C(t,\p)= \Theta(t)\rho(t,\p) +
G^<(t,\p).
\end{equation}
Using the ($t,\p$) representation and finally performing time Fourier
transfromation leads to
\begin{equation}
  S_{ab}(k) = \int\frac{d^3{\bf p}}{(2\pi)^3}\, \frac{dp_0}{2\pi}
  \,\frac{dp'_0}{2\pi} \,\frac{\rho_a(p_0,{\bf p}) \rho_b(p'_0,{\bf
  p}+{\bf k})}{k_0-p_0-p'_0+i\epsilon} \,(1+n(p_0) + n(p'_0)).
\label{kernelqu}
\end{equation}
Using free spectral functions $\rho_a(p)=(2\pi) \epsilon(p_0)
\delta(p^2-m_a^2)$ we arrive at the known expression 
(see for example \cite{boyanovsky96b})
\begin{eqnarray}
  S_{ab}(k) =&& \int\frac{d^3{\bf p}}{(2\pi)^3}
  \,\frac1{4\omega_a\omega_b}\,\biggl[  \frac{1+n_a+n_b}
  {k_0-\omega_a-\omega_b + i\epsilon
} - \frac{1+n_a+n_b}
  {k_0+\omega_a+\omega_b + i\epsilon} +\nonumber\\
  &&\frac{n_a-n_b} {k_0+\omega_a-\omega_b + i\epsilon} + \frac{n_b-n_a}
  {k_0-\omega_a+\omega_b + i\epsilon} \biggr].
\end{eqnarray}

On the other hand introducing in (\ref{kernelqu})
$\Delta^{(0)}(p)=G^>(p)=(1+ n(p_0))\rho(p)= G^<(-p)$, next performing
one of the $p_0$ integrals, and shifting properly the {\bf p} integral
we obtain
\begin{equation}
  S_{ab}(k) =\int\frac{d^4 p}{(2\pi)^4}\left[\frac{\Delta_{aa}^{(0)}(p)}
  {k^2+p^2-2kp -m_b^2} + \frac{\Delta_{bb}^{(0)}(-p)}{k^2+p^2  
  -2kp -m_a^2} \right].
\end{equation}
The $i\epsilon$ is assigned to $k_0$ by the Landau-prescription. Exploiting
that $\rho (p)\sim \delta(p^2-m^2)$  
one can write
\begin{equation}
  S_{ab}(k) =\int\frac{d^4 p}{(2\pi)^4}\left[\frac{\Delta_{aa}^{(0)}(p)}{k^2
  -2kp -M^2} + \frac{\Delta_{bb}^{(0)}(p)}{k^2 + 2kp +M^2} \right],
\end{equation}
where $M^2=m_b^2-m_a^2$. Finally performing a $\pm k/2$ shift the
result is
\begin{equation}
  S_{ab}(k) =\int\frac{d^4 p}{(2\pi)^4}\, \frac{\Delta_{bb}^{(0)}(p-k/2) -
  \Delta_{aa}^{(0)}(p +k/2)}{2kp + M^2}.
\label{sab_exp}
\end{equation}

Finally, introducing for the Fourier-transform of the left hand side
of Eq. (\ref{corr6eq}) the representation
\be
\langle\varphi_a\varphi_b\rangle=\int{d^4p\over (2\pi )^4}
\Delta^{(1)}(k,p),
\ee
we obtain from Eqs. (\ref{corr6eq}) and (\ref{sab_exp})
\begin{eqnarray}
  && 2kp \, \Delta_{aa}^{(1)}(k,p) = - \lambda_a \bar\Phi \Phi_1(k)
  \left[ \Delta_{aa}^{(0)}(p+k/2) - \Delta_{aa}^{(0)}(p-k/2)\right],
  \nonumber\\
  && (2kp +M_1^2) \, \Delta
_{1i}^{(1)}(k,p) = - \frac\lambda3 \bar\Phi
  \Phi_i(k) \left[ \Delta_{ii}^{(0)}(p+k/2) - 
  \Delta_{11}^{(0)}(p-k/2)\right],
\end{eqnarray}
which exactly coincides with Eq. (\ref{corr1eq}).

\section{Appendix}
\label{calss}
An alternative approach for calculating characteristic quantities of
real-time correlation functions is based on real-time dimensional
reduction \cite{nauta,jako1} and solving the resulting classical
effective theory relevant for the modes with high
occupation numbers. On-shell as well as the off-shell damping rates were
successfully reproduced \cite{jako2,aarts} in the symmetric phase of
$\phi^4$ theory using this approach. 
In this appendix we will discuss the application of
the classical approach to the $O(N)$ model and compare its results with the
exact one-loop dynamics.
The Lagrangian of the classical $O(N)$ theory has the following form:
\be
L_{cl}={1\over 2}{(\partial_{\mu}\tilde\pc_a)}^2-{1\over 2} m_{cl}^2(\Lambda)
{\tilde\pc_a}^2-{\lambda_{cl}\over 24}{(\tilde\pc_a^2)}^2-j_a\tilde\Phi_a.
\label{lclass}
\ee
The classical 
equation of motion corresponding to this Lagrangian is:
\be
(\partial^2+m_{cl}^2(\Lambda))\tilde \Phi_a +
{\lambda_{cl}\over 6} \tilde \Phi_a (\tilde \Phi_b^2)+j_a=0.
\ee
The external currents $j_a(x)$ were introduced in order to
prepare the derivation of the classical response theory.
They should not be confused with the induced currents $J_a$
appearing in the effectice quantum equations of motion.
The classical mass $m_{cl}(\Lambda)$ is different from
the mass parameter $m$ of the quantum theory (see Eq. (\ref{Lagr_dens})). 
The same is true for the coupling $\lambda_{cl}$.
The classical mass parameter depends on
the ultraviolet cutoff $\Lambda$. 
The results of the classical and quantum calculation
could be matched by a suitable choice of $m_{cl}(\Lambda)$ and
$\lambda_{cl}$.
In particular we shall see below (eq.(\ref{class_av})), that
the ultraviolet divergencies of the classical theory can be eliminated
if the divergent part of $m_{cl}(\Lambda)$ is suitably chosen
\cite{nauta,jako1,jako2,aarts}.
Therefore we will separate out the  divergent part from the classical
mass and write $m_{cl}^2(\Lambda)=m_T^2+\delta m^2(\Lambda)$
\footnote{The divergent part of the classical mass parameter will be treated 
as interaction, similarly to quantum field theory}.

In the broken symmetry phase one separates out the condensate
$\bar \Phi$,
\be
\tilde \Phi_a=\pv \delta_{a1}+\Phi_a
\ee
and the equations of motion read
\be
(\partial^2+m_T^2+{\lambda_{cl}\over2} \pv^2) \pc_1+{\lambda_{cl}\over 6} \pc_1
(\pc_a^2)+{\lambda_{cl}\over 2} \pv \pc_1^2 +{\lambda_{cl}\over 3}\pv
\pc_i^2+\delta m^2 \pc_1+j_1=0,
\label{cleq1}
\ee
\be
(\partial^2 + m_T^2 + {\lambda_{cl}\over 6} \pv^2)\pc_i+
{\lambda_{cl}\over 6} \pc_i (\pc_a^2) +{\lambda_{cl}\over
3}\pv \pc_1 \pc_i+\delta m^2 \pc_i+j_i=0.
\label{cleq2}
\ee
In addition the following initial conditions are imposed:
\be
\pc_a(t=0,\bx)=F_a(\bx), \qquad \partial_t \pc_a(t,\bx)|_{t=0}=P_a(\bx).
\ee
The expectation value of some quantity $O$ (e.g. some correlation 
function ) is obtained by averaging over the initial conditions
with the Boltzmann factor determined by the classical Hamiltonian
$H_{cl}(P_a,F_a,\pv)$ corresponding to (\ref{lclass})
\ba
&&
<O>={1\over Z}\int DF_a DP_a O \exp(-\beta H_{cl}(P_a,F_a,\pv))\\
&&
Z=\int DF_a DP_a \exp(-\beta H_{cl}(P_a,F_a,\pv)),
\ea
The explicit form of $H_{cl}(P_a,F_a,\pv)$ is
\ba
&&
H_{cl}(P_a,F_a,\pv)=\int d^3 x \biggl[
{1\over 2} P_a^2+{1\over 2}{(\partial_i F_a)}^2+
{1\over 2} (m_T^2 +{\lambda_{cl}\over 2}\pv^2)F_1^2+{1\over 2}(m_T^2+
{\lambda_{cl}\over 6} \pv^2)F_i^2+\nonumber\\
&&
{\lambda_{cl}\over 6} \pv F_1 {(F_b)}^2+
{\lambda_{cl}\over 24} {(F_a^2)}^2+(m_{cl}^2 \pv+{\lambda_{cl}\over
6}\pv^3) F_1+{1\over 2} m_{cl}^2 \pv^2+{\lambda_{cl}\over 24} \pv^4+
\delta m^2 F_a^2 \biggr],
\label{hclass}
\ea
where we have separated out the classical condansate $\pv$ since no
averaging over classical condensate is understood.
Since the classical condensate is separated both from the dynamical
fields and from the initial conditions the following equation holds:
\be
<\pc_1(t=0,\bx)>=<F_1(\bx)>=0.
\ee
Using the explicit form of $H_{cl}$ (Eq. (\ref{hclass}))
this results at one-loop level in the following 
equation for the classical condensate $\pv$:
\be
(m_{cl}^2(\Lambda)+{\lambda_{cl}\over 6} \pv^2)\pv+{\lambda_{cl}\over 6}\pv
\int {d^3 p\over {(2 \pi)}^3} \biggl[3{1\over
\bp^2+m_T^2+{\lambda_{cl}\over 2}\pv^2}+(N-1){1\over
\bp^2+m_T^2+{\lambda_{cl}\over 6}\pv^2}\biggr]=0.
\label{class_av}
\ee
Choosing $\delta m^2(\Lambda )$ to cancel the linearly divergent piece on
the left hand side this relation reduces in the high temperature limit
($m_T,\lambda_{cl} \pv\ll T$) to 
\be
m_T^2+{\lambda\over 6} \pv^2=0
\label{vac11}
\ee
and
\be
\delta m^2=-{\lambda_{cl}\over 6}(N+2){\Lambda T\over 2 \pi^2}.
\label{dm2}
\ee
Our procedure of solving the classical theory perturbatively closely follows
Ref. \cite{jako2}.
Equations
(\ref{cleq1}), (\ref{cleq2}) are rewritten in form of the following
integral equations:
\ba
&& 
\pc_1(x,j)=\pc_1^0(x)+\int d^4 x' D_R^1(x-x')({\lambda_{cl}\over 6} \pc_1
(\pc_a^2)+{\lambda_{cl}\over 2}\pv \pc_1^2+{\lambda_{cl}\over 3}\pv
\pc_i^2+\delta m^2 \pc_1+j_1)\nonumber\\
&&
\pc_i(x,j)=\pc_i^0(x)+\int d^4 x' D_R^i(x-x')({\lambda_{cl}\over 6} \pc_i
(\pc_a^2) +{\lambda_{cl}\over 3}\pv \pc_1 \pc_i+\delta m^2 \pc_i+j_i),
\label{inteq1}
\ea
where
\be
D_R^a(x-x')=-\theta(t) \int{d^3 q\over {(2 \pi)}^3}e^{i \bq
\bx}{\sin\omega_qt\over \w_q}
\ee
is the classical retarded Green function \cite{jako2} with
$\w_q=\sqrt{q^2+m_a^2}$, where $m_1^2=m_T^2+{\lambda_{cl}\over 2} \pv^2=
{\lambda_{cl}\over 3} \pv^2$ is the Higgs field mass and $m_i^2=m_T^2+
{\lambda_{cl}\over 6} \pv^2=0$ (we have used Eq. (\ref{vac11})).  
Furthermore, $\pc_a^0$ are the solutions of the free
equations of motion and $j$ in the argument of $\pc_a$ refers to the
functional dependence on $j=(j_1,j_i)$.  Following Ref. \cite{jako2}
one introduces linear response functions
\be
H_{R}^{ab}(x-x')={\delta \pc_a(x,j)\over\delta j_b(x')}.
\ee
Using the integral equation (\ref{inteq1}) one can derive a coupled set of
integral equations also
for the linear response functions $H_R^{ab}$ by functional differentiation
of eq. (\ref{inteq1})
(see Ref. \cite{jako2} for details). These integral equations can be
solved iteratively in the weak coupling limit. When solving the
equations, it is important to exploit the fact that only diagonal
components of $H_R^{ab}$ have terms ${\cal O}(\lambda^{0})$.  The
classical retarded response function is the ensemble average of
$H^{ab}_R$ with respect the initial conditions :
\be
G_{ab}^{cl}(x-x')=<H^{ab}_R(x-x')>.
\ee
By eq.(\ref{inteq1}) one easily finds that it satisfies
a Dyson-Schwinger equation of the general form
\be
G_{ab}^{cl}(x-x')=D_R^a(x-x')\delta_{ab}+
\int d^4y d^4 y' D_R^{a}(x-y) \delta_{ad}
\Pi^{cl}_{dc}(y-y') G_{cb}^{cl}
(y'-x'),
\label{ds}
\ee
where $\Pi_{ab}^{cl}(y-y')$ is the classical self-energy.

In the perturbative expansion the average is done with
the free Hamiltonian and therefore all thermal $n$-point functions are
expressed as products of the two-point function of the free fields
(solutions of the free equation of motion). 
The two point function of the free fields reads \cite{aarts, parisi} as
\be
\Delta_a^{cl,0}(x-x')=<\pc^0(x)_a \pc^0(x')_a>^0=
T \int {d^3 q\over {(2 \pi)}^3}e^{i \bq
(\bx-\bx')} {1\over \w_q^2} \cos(\w_q (t-t'))
\ee
This classical two point function is analogous to the free two point
function of the quantum theory $\Delta_{aa}^{(0)}(x,x')$
Using eq. (\ref{inteq1}) one gets the following  
self-energies for the Higgs and Goldstone fields at leading order in
the coupling constant $\lambda_{cl}$.
\ba
&&
\Pi_{11}^{cl}(\omega,\bk)=
\delta m^2+{\lambda_{cl}\over 6} \sum_a \Delta^{cl,0}_a(0,0)+
\nonumber\\
&&
{(\lambda_{cl} \pv)}^2 \int dt d^3y e^{i\w t-i \bk \by}\biggl(
\Delta_1^{cl,0}(y)D_R^1(y)+
{N-1\over 9} \Delta_i^{cl,0}(y) D_R^i(y)\biggr),
\nonumber\\
&&
\Pi_{ii}^{cl}(\w,\bk)=\delta m^2+
{\lambda_{cl}\over 6} \sum_a \Delta^{cl,0}_a(0,0)+\nonumber\\
&&
{({\lambda_{cl} \pv\over 3})}^2 \int dt d^3 y e^{i \w t-i \bk
\by} (\Delta^{cl,0}_1(y) D_R^i(y)+\Delta^{cl,0}_i(y) D_R^1(y) ),
\label{piclass}
\ea
Using the explicit form of $\Delta^{cl,0}_a(0,0)$ and Eq. (\ref{dm2})
one can easily verify that
all divergencies present in the above expression cancel.
Furthermore, in the high temperature limit and at leading order
in the coupling constant only the last terms contribute in the expression of
$\Pi_{11}^{cl}(\omega,\bk)$ and $\Pi_{ii}^{cl}(\omega,\bk)$.
One has to evaluate integrals of the following intrinsic form
\begin{equation}
  \label{kernelcl}
  \int dt \int d^3 y e^{i \w t -i \bk \by} D_R^a(y) \Delta^{cl,0}_{b}(y)
  =\int{d^3p\over {(2 \pi)}^3}\int {dp_0\over 2\pi} {dp_0'\over 2 \pi}
  {\rho_a(p_0,\w_a) \rho_b(p_0',\w_b)\over \w-p_0-p_0'+i \epsilon}
  (n^{cl}(p_0)+n^{cl}(p_0')),
\end{equation}
where $\w_a=\sqrt{\bp^2+m_a^2}$ and $\w_b=\sqrt{(\bp+\bk)^2+m_b^2}$.
The above expression coincides with the
result of the quantum calculation (\ref{kernelqu}), except the fact
the there is no $T=0$ contribution and the Bose-Einstein factors are
replaced by the classical distribution: $T/p_0$. Then it is easy to write down
the explicit expression for the classical self-energies using the
formal analogy with the result of the one-loop quantum calculations. 
For example for the
classical on-shell damping rate of the Goldstone modes one easily gets
the following expression:
\be
\Gamma_i^{cl}(\bk)=-{{\rm Im} \Pi^{cl}_{ii}(\w=|\bk|,\bk)\over 2 |\bk|}=
{\lambda_{cl}^2 \pv^2 \over 288 \pi^2 {| \bk |}^2} 
\int_{{m_{1}^2\over 4 |\bk|}}^{{m_{1}^2\over 4 |\bk|}+|\bk|}
dp n^{cl}(p)=
{\lambda_{cl}^2 \pv^2 T\over 288 \pi^2 {|\bk|}^2}
\ln\left(1+{4 {|\bk|}^2\over m_{1}^2}\right).
\label{giclass}
\ee

Now let us discuss the correspondence between classical and quantum 
calculations.
It was shown in Refs. \cite{nauta,jako1,jako2,aarts} that the result
of classical calculations can reproduce the high temperature
limit of the corresponding quantum results if the parameters
$m_{cl}^2(\Lambda)$ and $\lambda_{cl}$ are fixed to the values
determined
by static dimensional reduction. For our case this implies:
\be
m_{cl}^2(\Lambda)=m^2+{\lambda\over 6}(N+2)\biggl({T^2\over 12}-
{\Lambda T\over 2 \pi^2}\biggr),~~\lambda_{cl}=\lambda,
\ee
where $m$ and $\lambda$ are the mass and the coupling constant of the
corresponding quantum theory. From the above equations it is easy to 
see that the divergent part in $m_{cl}^2$ coincides with $\delta
m^2$ determined from the ultraviolet finitness of the classical result
(cf. Eqs. (\ref{class_av}),(\ref{dm2}) and (\ref{piclass})) and 
$m_1$ is the high temperature limit of $M_1$ (see Eqs. (\ref{vac11}) and
(\ref{avphi})).
For small values of $|\bk|$ ($|\bk| \ll m_{1}$) the logarithm in this
expression can be expanded and one obtains the result of
Eq. (\ref{goldstone_class}), which fails to reproduce the result of the
quantum calculation. The same is true for the whole imaginary part of
the Goldstone self-energy.  The high temperature limit of the
imaginary part of the Higgs self-energy (see Eq. (\ref{pis1})), on the
other hand, is well reproduced by the classical theory.

This result can be easily understood by looking at the effective
quantum equation of motion (\ref{effeqs}). In the equation for the
Goldstone fields the induced current $J_i$ has a
non-local contribution
from loop momenta $p\sim M_1^2/|\bk|>>T$ (cf. (\ref{impis})).
However, no such term is present in the corresponding classical
equation of motion on one hand and the classical theory cannot describe
fluctuation with wave length much smaller than $T^{-1}$ on the
other hand. The induced current $J_1$ in the effective equation of
motion for the Higgs fields receives non-local contribution only from
the loop momenta around $p \sim M_1$ which can be described in the
framework of the classical theory.

\section{Appendix}
\label{sec:appcalc}
In this Appendix we illustrate the steps of evaluation of the
integrals in Eq.~(\ref{pis}).

\paragraph{Imaginary parts.}

As example of the evaluation of a relevant integral we discuss in detail
 
\begin{equation}
  \textrm{Im} R_1(k,M)= \textrm{Im} \int \frac{d^4p}{(2\pi)^4}\,
  \frac{\Delta^{(0)}(p-k/2) - \Delta
^{(0)}(p+k/2)}{pk},
\end{equation}
where the propagators can be either of type $11$ or $ii$. In order to
implement the Landau-prescription we transform away any $k$ dependence
from the propagators by shifting the $p$ integral by $\pm k/2$. Then
we use
\begin{equation}
  \lim\limits_{\alpha\to 0}\textrm{Im}\frac1{x+i\alpha} = -\pi
  \epsilon(\alpha) \delta(x),
\end{equation}
and finally shift the integrals back. We write the 4D integration
measure as
\begin{equation}
  \int \frac{d^4p}{(2\pi)^4} = \frac1{8\pi^3}\int\limits_0^\infty\!
  dp\,p^2\!\int\limits_{-\infty}^\infty \!dp_0\!  \int\limits_{-1}^1
  \!dx,
\end{equation}
where $x=\hat{\bf p}\hat\k$ stands for the cosine of the angle between
the spatial momenta. The $x$ integration is trivial because it appears
in the Dirac-delta arising from the application of the principal value
theorem
\begin{equation}
  \int\limits_{-1}^1\!dx\,\delta(p_0k_0 - p|\k| x) = \frac1{p|\k|}\,
  \Theta(p|\k| - |p_0k_0|).
\end{equation}
Using the explicit form of the propagators
 (see
 Eq.~(\ref{propagator})) and the identity $\Theta(\omega)+n(|\omega|) =
\epsilon(\omega)(1+n(\omega))$ we find
\begin{equation}
  -\textrm{Im} R_1(k,M)=\frac1{4\pi|\k|}\int\limits_0^\infty\!dp\,p\!\!
  \int\limits_{-a}^a\!dp_0\, \epsilon(p_0\!-\!\frac{k_0}2)\,
  \epsilon(p_0\! +\! \frac{k_0}2)\, \delta(p_0^2\! -\! S^2) \left[
  n(p_0\!-\!\frac{k_0}2) - n(p_0\!+\!\frac{k_0}2)\right],
\end{equation}
where $a=p|\k|/|k_0|$ and $S^2=p^2+M^2-k^2/4$. The $p_0$ integration
over the Dirac-delta gives a constraint for the $p$ integration of the
form
\begin{equation}
  S<a \qquad\Rightarrow\qquad p^2\,\frac{k^2}{k_0^2} < \frac{k^2}4 -M^2.
\end{equation}
This can be fulfilled only for $k^2>4M^2$ (above the two-particle
threshold), or for $k^2<0$ (Landau damping). After elementary algebra
one can establish the value of the sign functions and one arrives at
the formula appearing in Eq.~(\ref{impis}).

A similar analysis can be performed for $\textrm{Im} R_i$, however, in
this case it proved to be more convenient to start from the equivalent
form
\begin{equation}
  \textrm{Im} R_i(k,M)= \textrm{Im} \int \frac{d^4p}{(2\pi)^4}\,
  \frac{\Delta_{11}^{(0)}(p-k) - \Delta_{ii}^{(0)}(p)}{2pk-k^2+M^2}.
\end{equation}
After implementing carefully the Landau prescription and performing
the $x$ integration as described before, we arrive at
\begin{equation}
  -\textrm{Im} R_i(k,M)=\frac1{8\pi|\k|}\int\limits_0^\infty\!dp\,p\!\!
  \int\limits_{b_-}^{b_+}\!dp_0\, \epsilon(p_0)\, \epsilon(p_0\!-\!k_0)\,
  \delta(p_0^2\! -\! p^2) \left[ n(p_0\!-\!k_0)-n(p_0)\right],
\label{gh_diff}
\end{equation}
where
\begin{equation}
  b_\pm = \frac{k^2-M^2}{2k_0} \pm p \frac{|\k|}{k_0}.
\end{equation}
The $p_0$ integration over the mass-shell delta-function again
restricts the domain of integration in the $p$ integral: $b_-<\pm
p<b_+$, which, however, does not restrict the possible values for
$k^2$. After the solution of these linear inequalities and the
analysis of the sign functions we arrive at the result appearing in
Eq.~(\ref{impis}).

\paragraph{Real parts.}

The relevant integrals in $\textrm{Re} R_1$ are
\begin{equation}
  I^{\pm} = \textrm{Re} \int\!\frac{d^4p}{(2\pi)^4}\,
  \frac{\Delta^{(0)}(p)}{pk \pm k^2/2},
\end{equation}
where the propagator can be either of type $11$ or $ii$. With their
help we find $\textrm{Re} R_1 =I^+ - I^-$. The real part comes from
the principal value integration. We decompose the integration measure
as in the calculation of the imaginary parts. When we use
Eq.~(\ref{propagator}) for the propagator, the value of $p_0$ is fixed
by the delta function. The $x$ integration can be performed as
\begin{equation}
  \int\limits_{-1}^1\!dx\,{\cal P}\frac1{2p_0k_0 -2p|\k| x \pm k^2} =
  \frac1{p|\k|} \textrm{arth}(\frac{2p|\k|}{2p_0k_0 \pm k^2}),
\end{equation}
where $\textrm{arth}(x) =1/2 \ln|(1+x)/(1-x)|$. Using the
properties of the absolute value we find 
\begin{equation}
  I^+(k_0,\k) = - I^{-}(-k_0,\k).
\end{equation}
Then we directly arrive at Eq.~(\ref{repis}).

We write $\textrm{Re} R_i$ in the form
\begin{equation}
  \textrm{Re} R_i = \textrm{Re} \int\!\frac{d^4p}{(2\pi)^4}\,
  \frac{\Delta_{11}^{(0)}(p)}{2pk + k^2 +M^2} - \textrm{Re}
  \int\!\frac{d^4p}{(2\pi)^4}\, \frac{\Delta_{ii}^{(0)}(p)}{2pk - k^2
  +M^2}.
\end{equation}
Then the previous scheme of calculation goes through directly.

The only problem still to be discussed is the zero temperature
contribution, which diverges logarithmically. The regularization and
renormalization prescriptions are better formalized in the language of
the propagators and couplings. Using the results of
Appendix~\ref{app:pertth} to go over to the perturbation theory, we
have to evaluate
\begin{equation}
  R=\int\frac{d^4p}{(2\pi)^4}\, \left[ G^C(p) G^C(p-k) -
    G^<(p) G^<(p-k) \right].
\label{req}
\end{equation}
The first term at $T=0$ is the usual time ordered product;
with finite four-dimensional cutoff one finds
\begin{equation}
  \int\frac{d^4p}{(2\pi)^4}\,G^C(p) G^C(p-k) =
  \frac{-1}{16\pi^2} \left[ 1 + \int\limits_0^1\!dx\, \ln\frac{|k^2
  x(1-x) -m_1^2 x -m_2^2 (1-x)|}{\Lambda^2} \right].
\end{equation}
The divergence is canceled by the coupling constant counterterm. In
modified minimal subtraction ($\overline{\textrm{MS}}$) scheme it reads
\begin{equation}
  \frac{-1}{16\pi^2}\int_0^1dx \ln\frac{|k^2 x(1-x) -m_1^2 x -m_2^2
  (1-x)|}{\mu^2}.
\end{equation}

In the second term of Eq. (\ref{req}) at $T=0$ we can use
$G^<(p)=\Theta(-p_0)(2\pi) \delta(p^2-m^2)$. Because of the delta
functions this piece yields finite result (as is expected by the
arguments of the renormalizability)
\begin{equation}
  \int\frac{d^4p}{(2\pi)^4}\,G^<(p) G^<(p-k) = \frac1{8\pi} \left\{
  \begin{array}[c]{ll}
    \sqrt{1-\frac{4m^2}{k^2}}\,\Theta(k^2-4m^2), & \textrm{if}\,
    ~m_1=m_2=m \cr (1-\frac{m^2}{k^2})\,\Theta(k^2-m^2), &
    \textrm{if}\, ~m_1=0,\,m_2=m. \cr
  \end{array}\right.
\end{equation}

\end{appendix}
\section*{Acknowledgements}
The authors thank D. Boyanovsky and H.J. de Vega for valuable 
correspondence on the subject of
 this paper. Financial support of OTKA
(Hungarian Science Fund) is gratefully acknowledged.

\end{document}